# Quantum Mechanical Potentials and Inactive Electron Effects in Resonant Charge Exchange Collisions

A. L. Harris and A. Plumadore

Physics Department, Illinois State University, Normal, IL, USA 61790

## Abstract

Scattering angle differential cross sections for the $He^+ + He$ single electron capture process are studied using plane wave Born approximation models for projectile energies between 30 keV and 1.89 MeV. Within this simplistic framework, we study the effects of the frozen core approximation by performing a full 5-Body calculation that explicitly includes all particles in the collision and comparing it with a single bound state model that neglects the bound electron in the projectile and a double bound state model that neglects the inactive electron in the target atom. Results are compared with experiment and we show that inclusion of the inactive electron in the perturbation potential is more important than inclusion in the wave functions. We also introduce a semi-quantum mechanical perturbation potential that treats the atomic electrons as a quantum mechanical electron cloud rather than point particles. The semi-quantum mechanical perturbation removes the deep, unphysical minimum that exists in cross sections calculated with Born-type models, but also has the effect of greatly reducing the magnitude of the small scattering angle cross sections.

## 1. Introduction

Electron capture processes play an important role in many physical systems, from fusion reactors to astrophysical processes. In an electron capture collision, an incident ion collides with a target atom, captures an electron, and leaves the collision as a bound state. Recent experimental results for heavy ion single electron capture at low projectile velocities exhibit features in the differential cross section that can be attributed to Fraunhofer diffraction effects [1-6]. At high velocities, the two-step Thomas mechanism becomes important in order to accurately describe the differential cross sections [7-9]. In the intermediate energy range, recent work has shown that projectile coherence effects can be important, especially when the collision target is a small molecule, such as $H_2$ [10-14]. In electron capture collisions, it is common to neglect inactive electrons in what is known as the frozen core approximation. Some of our previous work has focused on the validity of the frozen core approximation in electron and heavy ion ionization collisions [15,16]. Here, we extend this work to electron capture differential cross sections with dressed projectiles.

    We introduce several Plane Wave Born Approximation (PWBA) models that are used to explore the effects of neglecting inactive electrons in the collision. We also introduce variants of these models that treat the interaction potential between the projectile and target atom semi-quantum mechanically rather than classically. All of the models are then compared to the experimental results of [1] and [17]. For the sake of clarity, and to make comparisons with recent experiments, we focus exclusively on the $He^+$ + He single electron capture process with all bound electrons in the ground state. The experimental data sets with which we compare our models use both $^3He^+$ and $^4He^+$ isotopes as projectiles. From an experimental standpoint, the distinguishability of the projectile from the target is a clear advantage. The effect of the isotope on the theoretical models is minimal and only appears in the mass of the projectile. There is no effect on the charges of the particles in the collision, and any effect on binding energies is so small it can be safely neglected.

The PWBA is the simplest perturbative model available and is well-known to overestimate the magnitude of the scattering cross sections. However, the use of such a simplistic model still holds value in qualitatively studying the physical effects in which we are interested, particularly at larger projectile velocities. The models presented here should be viewed as a guide for assessing the relative importance of certain physical effects and hopefully influence the development of more sophisticated models. Atomic units are used throughout unless otherwise stated.

## 2. Theory

We present results for angular differential cross sections from three different fully quantum mechanical Plane Wave Born Approximation (PWBA) models. In all models, the center of mass cross section is given by

$$\frac{d\sigma^C}{d\Omega} = \frac{(2\pi)^4 \mu_{pa} \mu_{pi} k_f}{k_i} |T_{fi}|^2 , \tag{1}$$

where $\mu_{pa}$ is the reduced mass of the initial state projectile and target atom, $\mu_{pi}$ is the reduced mass of the scattered projectile and residual ion, $k_f(k_i)$ is the center of mass momentum of the scattered (incident) projectile, and $T_{fi}$ is the transition matrix given by

$$T_{fi} = <\Psi_f|V_i|\Psi_i> . \tag{2}$$

The lab frame cross section is related to the center of mass cross section by

$$\frac{d\sigma^L}{d\Omega} = \left[\frac{(1+2\delta \cos\theta_c + \delta^2)^{\frac{3}{2}}}{|1+\delta \cos\theta_c|}\right] \frac{d\sigma^C}{d\Omega} , \tag{3}$$

where $\theta_C$ is the scattering angle of the projectile in the center of mass frame. The quantity $\delta$ is given by the ratio of the magnitude of the velocity of the center of mass of the entire collision system in the lab frame $V$ and the magnitude of the scattered projectile velocity in the center of mass frame $v_f^C$, such that $\delta = \frac{V}{v_f^C}$. The lab frame and center of mass frame scattering angles are related by

$$\tan\theta_L = \frac{\sin\theta_C}{\cos\theta_C + \delta}. \tag{4}$$

The form of the initial (final) wave function $\Psi_i(\Psi_f)$, and the perturbation $V_i$ are dependent upon the model used, although in each case, we assume independent electrons in the bound states. Then, two-particle bound states are represented by a simple product variational wave function

$$\Phi_{He}(\vec{x},\vec{y}) = \frac{\alpha^3}{\pi} e^{-\alpha x} e^{-\alpha y} \tag{5}$$

with $\alpha = 1.6875$ as the screening parameter. This choice of independent electron helium atom wave function neglects electron correlation within the atom, but it has been shown that for electron capture without excitation, electron correlation plays a minimal role in the cross sections [17,18]. In all models, the motion of the initial and final state continuum projectiles are treated as plane waves and bound states involving only one active electron are purely hydrogenic. The 1s hydrogenic wave function with nuclear charge Z is given by

$$\phi(\vec{x}) = \frac{Z^{3/2}}{\sqrt{\pi}} e^{-Zx}. \tag{6}$$

Additionally, for the analytical calculations it is necessary to use the momentum space wave function for a 1s hydrogen-like atom with nuclear charge Z given by

$$\phi^{FT}(\vec{p}) = \frac{1}{(2\pi)^{3/2}} \int e^{i\vec{p}\cdot\vec{x}} \phi(\vec{x}) d\vec{x} = \frac{4Z^{5/2}}{\pi\sqrt{2}(Z^2+p^2)^2}. \tag{7}$$

Equation (2) can then be written as an integral over the spatial coordinates of all particles included in the calculation. Because of the approximations made, most of the integrations can be performed analytically, with only a 3-dimensional integral that is performed numerically.

Below is a description of each of the models and Table 1 includes a comparison of the wave functions and perturbations used in Eq. (2). Calculations are performed in the center of mass frame using Jacobi coordinates, however results are presented in the lab frame for comparison with experiment. The diagrams in Figs 1 through 4 show the coordinate systems used for each of the models. We note that the diagrams and equations used here describe distinguishable electrons, but our calculations properly account for the indistinguishability of the electrons in the target atom. However, in the case of resonant charge transfer we neglect exchange of the alpha particles.

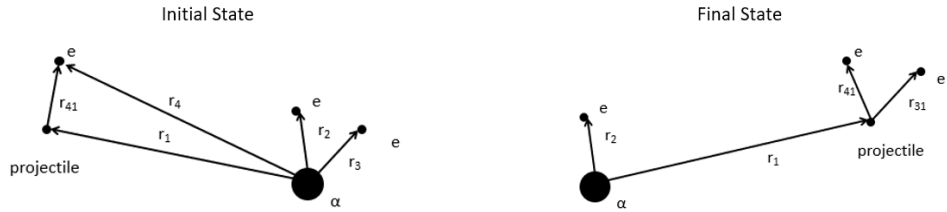

**Figure 1 Lab frame coordinates for a collision between a projectile and helium atom.**

**2.1 5-Body Model**

The He$^+$ + He collision system is inherently a 5-body system, consisting of three electrons and two nuclei. Within the PWBA and independent electron model, it is possible to perform a complete 5-body calculation that includes all particles in the collision system and does not invoke the frozen core approximation. In this case, the incident projectile bound state is purely hydrogenic (Eq. (6)), while the target atom is a bound state represented by the product of two hydrogen-like orbitals (Eq. (5)). The initial state wave function is then given by

$$\Psi_i = \chi_i(\vec{R_i})\phi_p(\vec{r_{41}})\Phi_{He}(\vec{r_2},\vec{r_3}), \qquad (8)$$

where $\chi_i(\vec{R_i})$ is the incident projectile plane wave, $\phi_p(\vec{r_{41}})$ is the incident projectile bound state, and $\Phi_{He}(\vec{r_2},\vec{r_3})$ is the target atom wave function.

After the collision, the scattered projectile is a bound state represented by the product of two hydrogen-like orbitals (Eq. (5)), while the residual ion is purely hydrogenic (Eq. (6)). The final state wave function is given by

$$\Psi_f = \chi_f(\vec{R_f})\Phi_{pb}(\vec{r_{41}}, \vec{r_{31}})\xi_{He^+}(\vec{r_2}), \tag{9}$$

where $\chi_f(\vec{R_f})$ is the scattered projectile plane wave, $\phi_{pb}(\vec{r_{41}}, \vec{r_{31}})$ is the captured electron bound state, and $\xi_{He^+}(\vec{r_2})$ is the residual ion wave function.

In a traditional PWBA model, the perturbation includes all two particle Coulomb interaction terms between the incident projectile and the target atom and is given by

$$V_i = \frac{Z_\alpha Z_\alpha}{r_1} + \frac{Z_\alpha Z_e}{r_{12}} + \frac{Z_\alpha Z_e}{r_{13}} + \frac{Z_e Z_e}{r_{43}} + \frac{Z_e Z_e}{r_{42}} + \frac{Z_e Z_\alpha}{r_4}, \tag{10}$$

where $Z_e$ is the charge of the electron and $Z_\alpha$ is the charge of the alpha particle. In addition to this classical perturbation, we also consider a perturbation in which the nucleus is a point particle but the electron charge cloud is treated quantum mechanically. The potential generated by the atom is then

$$V_{atom}(r) = \frac{Z_\alpha}{r} + V_e(r), \tag{11}$$

where $V_e(r)$ is the potential of the atomic electron cloud. It can be found by integrating the electronic wave function to get the charge density, using Gauss's law to find the electric field, and then calculating the potential. In the 5-Body model, this semi-quantum mechanical (SQM) perturbation is given by

$$V_i = Z_\alpha V_{atom}(r_1) + Z_e V_{atom}(r_4). \tag{12}$$

For the variational wave function of Eq. (5), the target atom electronic potential is

$$V_e(r) = 2\alpha \left( e^{-2\alpha r} + \frac{e^{-2\alpha r}}{\alpha r} - \frac{1}{\alpha r} \right). \tag{13}$$

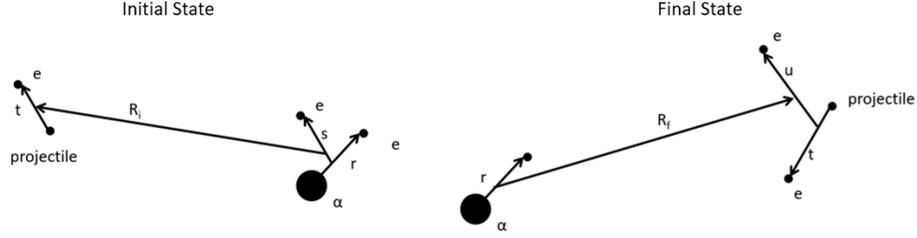

**Figure 2 Initial and final state Jacobi coordinates for the 5-Body model.**

Jacobi coordinates are used to calculate the center of mass T-matrix and are shown in Fig 2. They are related to the lab coordinates of Fig. 1 by

$$\vec{R_i} = \frac{m_p \vec{r_1}}{m_e + m_p} - \frac{m_e(\vec{r_2} + \vec{r_3})}{2m_e + m_\alpha} + \frac{m_e \vec{r_4}}{m_e + m_p} \tag{14}$$

$$\vec{R_f} = \frac{m_p \vec{r_1}}{2m_e + m_p} + \frac{m_e(\vec{r_3} + \vec{r_4})}{2m_e + m_p} - \frac{m_e \vec{r_2}}{m_e + m_\alpha} \tag{15}$$

$$\vec{s} = \vec{r_3} - \frac{m_e \vec{r_2}}{m_e + m_\alpha} \tag{16}$$

$$\vec{r} = \vec{r_2} \tag{17}$$

$$\vec{t} = \vec{r_{41}} \tag{18}$$

$$\vec{u} = \vec{r_3} - \frac{m_p \vec{r_1}}{m_e + m_p} - \frac{m_e \vec{r_4}}{m_e + m_p} \tag{19}$$

where $m_e$ is the mass of the electron, $m_\alpha$ is the mass of the alpha particle nucleus and $m_p$ is the mass of the projectile nucleus. Note that in the initial state target atom wave function we have assumed $\vec{s} \approx \vec{r_3}$ and in the final state scattered projectile wave function, we have assumed $\vec{u} = \vec{r_{31}}$.

Inserting the wave functions from Eqs (8) and (9) along with the classical perturbation of Eq (10) into Eq. (2) yields the 5-Body model transition matrix

$$T_{fi}^{cl} = \frac{256\alpha^8 Z_\alpha^3 (Z_\alpha+\alpha)^{\frac{1}{2}}}{\pi^4} \int \frac{d\vec{p}}{(\alpha^2+p^2)^2 |\vec{A}+\vec{D}-\vec{p}|^2} \left( \frac{Z_\alpha}{[(Z_\alpha+\alpha)^2+D^2]^2} + \right.$$

$$\left. \frac{Z_e}{\left[(Z_\alpha+\alpha)^2+|\vec{A}-\vec{p}|^2\right]^2} \right) \left( \frac{1}{\left[\alpha^2+|\vec{p}-\vec{C}|^2\right]^2} \left( \frac{Z_e}{\left[(Z_\alpha+\alpha)^2+|\vec{A}+\vec{D}-\vec{p}-\vec{B}|^2\right]^2} + \frac{Z_\alpha}{[(Z_\alpha+\alpha)^2+B^2]^2} \right) + \right.$$

$$\left. \frac{Z_e}{\left[\alpha^2+|\vec{A}+\vec{D}-\vec{C}|^2\right]^2 [(Z_\alpha+\alpha)^2+B^2]^2} \right), \tag{20}$$

where $\vec{A}, \vec{B}, \vec{C}$ and $\vec{D}$ are related to the center of mass momenta $\vec{k_i}$ and $\vec{k_f}$ by

$$\vec{A} = \frac{m_p \vec{k_i}}{m_e+m_p} - \frac{m_p \vec{k_f}}{2m_e+m_p} \tag{21}$$

$$\vec{B} = \frac{m_e \vec{k_i}}{2m_e+m_\alpha} - \frac{m_e \vec{k_f}}{m_e+m_\alpha} \tag{22}$$

$$\vec{C} = \frac{m_e \vec{k_i}}{2m_e+m_\alpha} - \frac{m_e \vec{k_f}}{2m_e+m_p} \tag{23}$$

$$\vec{D} = \frac{m_e \vec{k_i}}{m_e+m_p} - \frac{m_e \vec{k_f}}{2m_e+m_p} \tag{24}$$

The semi-quantum mechanical perturbation still permits a mostly analytical calculation with the 5-Body SQM transition matrix given by

$$T_{fi}^{SQM} = \frac{256\alpha^{\frac{13}{2}} Z_\alpha^3 (Z_\alpha+\alpha)^2}{\pi^2 [(Z_\alpha+\alpha)^2+B^2]^2} \int d\vec{p} \frac{\left(8\alpha^2+|\vec{A}+\vec{D}-\vec{p}|^2\right)}{[\alpha^2+p^2]^2 \left[\alpha^2+|\vec{p}-\vec{C}|^2\right]^2 \left[4\alpha^2+|\vec{A}-\vec{p}+\vec{D}|^2\right]^2} \left( \frac{Z_\alpha}{[(Z_\alpha+\alpha)^2+D^2]^2} + \right.$$

$$\left. \frac{Z_e}{\left[(Z_\alpha+\alpha)^2+|\vec{A}-\vec{p}|^2\right]^2} \right). \tag{25}$$

The three-dimensional integrals over $\vec{p}$ are performed numerically using Gauss-Legendre quadrature.

### 2.2 Single Bound State Model

In the single bound state (SBS) model, the incident projectile is treated using the frozen core approximation and is assumed to be structureless with a charge $Z_{pi}$ and mass $m_e + m_p$. This model neglects any bound state effects of the incident He$^+$ ion and is identical to that used in [19], except for the difference in mass of the projectile. The target helium atom is a two-electron atom. In this model, the initial state wave function is given by

$$\Psi_i = \chi_i(\vec{R_i})\Phi_{He}(\vec{r_2},\vec{r_3}), \qquad (26)$$

where $\chi_i(\vec{R_i})$ is the incident projectile plane wave and $\Phi_{He}(\vec{r_2},\vec{r_3})$ is the target atom wave function. The final state consists of a scattered one-electron bound state projectile and one-electron residual helium ion. The final state wave function is then given by

$$\Psi_f = \chi_f(\vec{R_f})\phi_p(\vec{r_{31}})\xi_{He^+}(\vec{r_2}), \qquad (27)$$

where $\chi_f(\vec{R_f})$ is the scattered projectile plane wave, $\phi_p(\vec{r_{31}})$ is the captured electron bound state, and $\xi_{He^+}(\vec{r_2})$ is the residual ion wave function. We again consider classical and semi-quantum mechanical perturbations. The classical perturbation consists of all two-particle interaction terms between the projectile and the two-electron helium target and in the SBS model it is given by

$$V_i = \frac{Z_{pi}Z_\alpha}{r_1} + \frac{Z_{pi}Z_e}{r_{12}} + \frac{Z_{pi}Z_e}{r_{13}}, \qquad (28)$$

where $Z_{pi}$ is the charge of the incident projectile. The SQM perturbation in the SBS model is

$$V_i = Z_{pi}V_{atom}(r_1), \qquad (29)$$

where $V_{atom}$ is given by Eq. (11). The Jacobi coordinates for the SBS model are shown in Fig 3. Note that because the projectile is assumed structureless, there is no $\vec{r_4}$ coordinate in the SBS model.

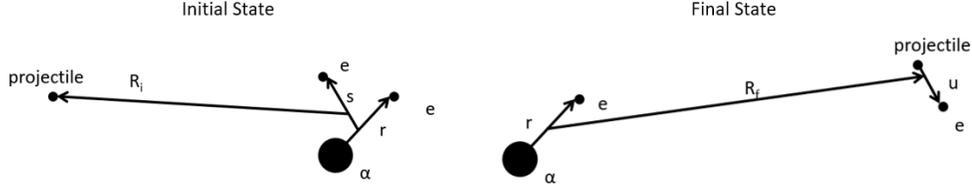

**Figure 3 Initial and final state Jacobi coordinates for the single bound state model.**

The relationship between the lab frame coordinates and the Jacobi coordinates are given by

$$\vec{R_i} = \vec{r_1} - \frac{m_e(\vec{r_2}+\vec{r_3})}{2m_e+m_\alpha} \tag{30}$$

$$\vec{R_f} = \frac{(m_e+m_p)\vec{r_1}}{m_e+m_p} + \frac{m_e\vec{r_3}}{m_e+m_p} - \frac{m_e\vec{r_2}}{m_e+m_\alpha} \tag{31}$$

$$\vec{s} = \vec{r_3} - \frac{m_e\vec{r_2}}{m_e+m_\alpha} \tag{32}$$

$$\vec{r} = \vec{r_2} \tag{33}$$

$$\vec{u} = \vec{r_{31}} \tag{34}$$

Note that again we have assumed that $\vec{s} \approx \vec{r_3}$. Inserting all of the wave function expressions and the two choices of perturbation for the SBS model into Eq (2) yields the following expressions for the transition matrices

$$T_{fi}^{cl} = \frac{8Z_\alpha^{3/2} Z_{pi} Z_{pf}^{5/2} \alpha^4 (Z_\alpha+\alpha)}{\pi^{7/2}} \int \frac{d\vec{p}}{\left(Z_{pf}^2+p^2\right)^2 |\vec{A}-\vec{p}|^2} \left[ \frac{Z_\alpha}{\left(\alpha^2+|\vec{p}-\vec{B}|^2\right)^2 (C^2+(\alpha+Z_\alpha)^2)^2} + \right.$$

$$\left. \frac{Z_e}{\left(\alpha^2+|\vec{A}-\vec{B}|^2\right)^2 (C^2+(\alpha+Z_\alpha)^2)^2} + \frac{Z_e}{\left((Z_\alpha+\alpha)^2+|\vec{C}-\vec{A}+\vec{p}|^2\right)^2 \left(\alpha^2+|\vec{p}-\vec{B}|^2\right)^2} \right], \tag{35}$$

and

$$T_{fi}^{SQM} = \frac{64\alpha^4 (Z_\alpha+\alpha) Z_\alpha^{\frac{3}{2}} Z_{pi} Z_{pf}^{\frac{5}{2}}}{\pi^4 [(Z_\alpha+\alpha)^2+C^2]^2} \int d\vec{p} \frac{\left(8\alpha^2+|\vec{A}-\vec{p}|^2\right)}{\left[Z_{pf}^2+p^2\right]^2 \left[4\alpha^2+|\vec{A}-\vec{p}|^2\right]^2 \left[\alpha^2+|\vec{p}-\vec{B}|^2\right]^2}, \tag{36}$$

where now $\vec{A} = \vec{k_i} - \frac{m_p \vec{k_f}}{m_e+m_p}$, $\vec{B} = \frac{m_e \vec{k_i}}{m_\alpha+2m_e} + \frac{m_e \vec{k_f}}{m_e+m_\alpha}$, and $\vec{C} = \frac{m_e \vec{k_i}}{m_\alpha+2m_e} + \frac{m_e \vec{k_f}}{m_e+m_p}$.

## 2.3 Double Bound State Model

In the Double Bound State (DBS) model, the electronic structure of the incident projectile is now included in the calculation. However, the target atom wave function is treated with the frozen core approximation and therefore modeled as a single active electron wave function for effective charge $Z_{ti}$ and mass $m_t = m_e + m_\alpha$. This results in the calculation not including the coordinate $\vec{r_2}$. The initial state wave function is given by

$$\Psi_i = \chi_i(\vec{R_i})\Phi_{He}(\vec{r_3})\phi_p(\vec{r_{41}}), \tag{37}$$

where $\chi_i(\vec{R_i})$ is the incident projectile plane wave, $\Phi_{He}(\vec{r_3})$ is the target atom wave function, and $\phi_p(\vec{r_{41}})$ is the incident projectile bound state electron wave function. In the final state, the scattered projectile is a two-electron helium atom and the residual ion is a structureless point particle of charge $Z_{ti}$. The final state wave function is given by

$$\Psi_f = \chi_f(\vec{R_f})\Phi_{pb}(\vec{r_{41}},\vec{r_{31}}), \tag{38}$$

where $\chi_f(\vec{R_f})$ is the scattered projectile plane wave and $\Phi_{pb}(\vec{r_{41}},\vec{r_{31}})$ is the scattered projectile bound state wave function. The classical perturbation is given by

$$V_i = \frac{Z_\alpha Z_{ti}}{r_1} + \frac{Z_\alpha Z_e}{r_{13}} + \frac{Z_e Z_e}{r_{43}} + \frac{Z_e Z_{ti}}{r_4}. \tag{39}$$

The SQM perturbation is now

$$V_i = Z_\alpha V_{atom}(r_1) + Z_e V_{atom}(r_4) \tag{40}$$

with $V_{atom}(r)$ found in a similar manner to that of Eq. (11) using charge density and Gauss's law. However, because the target atom is a one-electron hydrogenic wave function, the charge appearing in the potential is different and the result is smaller by a factor of ½. This results in

$$V_e(r) = Z_{ti}\left[e^{-2Z_{ti}r} + \frac{e^{-2Z_{ti}r}}{Z_{ti}r} - \frac{1}{Z_{ti}r}\right]. \tag{41}$$

The Jacobi coordinates used for the DBS model are shown in Fig 4.

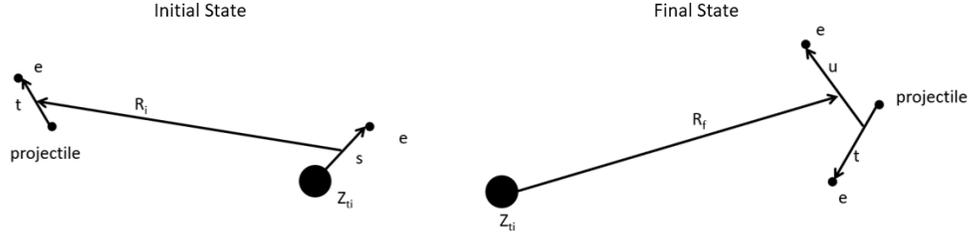

**Figure 4 Initial and final state Jacobi coordinates for the double bound state model.**

The expressions linking the lab frame coordinates to the Jacobi coordinates are

$$\vec{R_i} = \frac{m_p \vec{r_1}}{m_e+m_p} - \frac{m_e \vec{r_3}}{2m_e+m_\alpha} + \frac{m_e \vec{r_4}}{m_e+m_p} \tag{42}$$

$$\vec{R_f} = \frac{m_p \vec{r_1}}{2m_e+m_p} + \frac{m_e(\vec{r_3}+\vec{r_4})}{2m_e+m_p} \tag{43}$$

$$\vec{u} = \vec{r_3} - \frac{m_p \vec{r_1}}{m_e+m_p} - \frac{m_e \vec{r_4}}{m_e+m_p} \tag{44}$$

$$\vec{t} = \vec{r_{41}} \tag{45}$$

$$\vec{s} = \vec{r_3} \tag{46}$$

As before, we make the approximation that $\vec{u} \approx \vec{r_{31}}$.

Inserting all of the wave functions and the two choices of perturbation for the DBS model into Eq (2) yields

$$T_{fi}^{cl} = \frac{2(Z_\alpha+\alpha)\alpha^4 Z_{ti}^{5/2} Z_\alpha^{3/2}}{\pi^7} \int \frac{d\vec{p}}{(\alpha^2+p^2)^2 |\vec{D}-\vec{p}-\vec{G}|^2} \left( \frac{Z_e}{\left(Z_{ti}^2+|\vec{D}-\vec{G}-\vec{F}|^2\right)^2} + \frac{Z_{ti}}{\left(Z_{ti}^2+|\vec{p}-\vec{F}|^2\right)^2} \right) \left( \frac{Z_\alpha}{((\alpha+Z_\alpha)^2+G^2)^2} + \frac{Z_e}{\left((Z_\alpha+\alpha)^2+|\vec{D}-\vec{p}|^2\right)^2} \right) \tag{47}$$

and

$$T_{fi}^{SQM} = \frac{32\alpha^4 Z_\alpha^{\frac{3}{2}} Z_{ti}^{\frac{5}{2}}(Z_\alpha+\alpha)}{\pi^4} \int d\vec{p} \frac{1}{[\alpha^2+p^2]\left[Z_{ti}^2+|\vec{p}-\vec{F}|^2\right]^2} \left( \frac{Z_\alpha}{[(Z_\alpha+\alpha)^2+G^2]^2} + \right.$$

$$\left. \frac{Z_e}{\left[(Z_\alpha+\alpha)^2+|\vec{D}-\vec{p}|^2\right]^2} \right) \left( \frac{8Z_{ti}^2+|\vec{D}-\vec{p}-\vec{G}|^2}{\left[4Z_{ti}^2+|\vec{D}-\vec{p}-\vec{G}|^2\right]^2} + \frac{Z_{ti}-1}{|\vec{D}-\vec{p}-\vec{G}|^2} \right), \tag{48}$$

where $\vec{D} = \frac{m_p \vec{k}_i}{m_e+m_p} - \frac{m_p \vec{k}_f}{2m_e+m_p}$, $\vec{F} = \frac{m_e \vec{k}_i}{2m_e+m_\alpha} + \frac{m_e \vec{k}_f}{2m_e+m_p}$, and $\vec{G} = \frac{m_e \vec{k}_i}{m_e+m_p} - \frac{m_e \vec{k}_f}{2m_e+m_p}$.

Finally, we introduce one additional variant of the DBS model in which the potential of the target atom frozen core is modeled with a screened Coulomb (Yukawa) potential. In this case, the classical perturbation of the DBS model is modified such that the two-particle interaction between the projectile constituents and the target nucleus is multiplied by a screening term, with screening parameter $\beta$. In this DBS-Yukawa (DBS-Y) model, the perturbation is

$$V_i = \frac{Z_\alpha Z_{ti} e^{-\beta r_1}}{r_1} + \frac{Z_\alpha Z_e}{r_{13}} + \frac{Z_e Z_e}{r_{43}} + \frac{Z_e Z_{ti} e^{-\beta r_4}}{r_4}. \tag{49}$$

If $\beta = 0$, this perturbation reduces to that of the classical DBS perturbation. The DBS-Y transition matrix is given by

$$T_{fi}^Y = \frac{32\alpha^{\frac{3}{2}} Z_\alpha^{\frac{3}{2}} Z_{ti}^{\frac{5}{2}}(Z_\alpha+\alpha)}{\pi^4} \int d\vec{p} \frac{1}{[\alpha^2+p^2]^2} \left( \frac{Z_{ti}}{\left[\beta^2+|\vec{D}-\vec{p}-\vec{G}|^2\right]^2 \left[Z_{ti}^2+|\vec{p}-\vec{F}|^2\right]^2} + \right.$$

$$\left. \frac{Z_e}{|\vec{D}-\vec{p}-\vec{G}|^2 \left[Z_{ti}^2+|\vec{D}-\vec{p}-\vec{G}|^2\right]^2} \right) \left( \frac{Z_\alpha}{[(Z_\alpha+\alpha)^2+G^2]^2} + \frac{Z_e}{\left[(Z_\alpha+\alpha)^2+|\vec{D}-\vec{p}|^2\right]^2} \right) \tag{50}$$

For a comparison of the classical, SQM, and Yukawa perturbations used, Fig 5 shows $V_{atom}$ for the SBS-SQM, DBS-SQM, Coulomb, and Yukawa potentials, as well as the charge enclosed $Q(r)$ as a function of radial distance. Note that the Coulomb and Yukawa potentials shown are those of the target atom nucleus only and do not include the electronic potential. However, the curves for $V_{atom}$ include both the electronic and nuclear potentials. From Fig 5, it can be seen that other than the Coulomb nuclear potential, the Yukawa nuclear potential decays most slowly.

This results in the Yukawa nuclear charge distribution being the broadest. The SMS SQM charge enclosed is two at the origin, while the DBS SQM charge enclosed is one at the origin. This difference is due to the use of the frozen core approximation in the target atom wave function in the DBS model, and leads to the DBS charge enclosed having a greater range than the SBS charge.

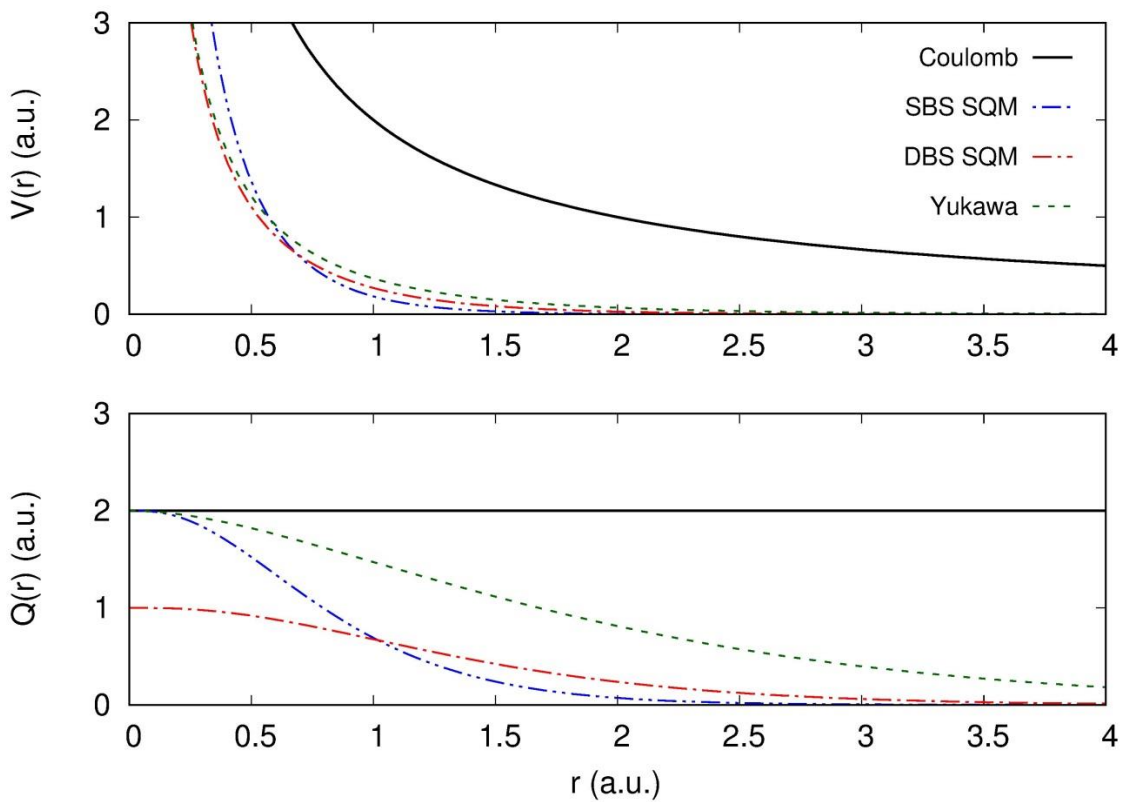

**Figure 5 Model potentials used in the classical, SQM, and Yukawa perturbations. The top panel shows the potential vs radial distance from the nucleus and the bottom panel shows charge enclosed in a sphere of radius r. The parameters used were $\alpha = 1.6875, \beta = 1$, $Z_{pi} = 1$, and $Z_{ti} = 1$. For the Coulomb potential $V(r) = \frac{2}{r}$ and for the Yukawa potential $V(r) = \frac{2}{r} e^{-\beta r}$. Note that the Coulomb and Yukawa potentials shown are those of the target atom nucleus only and do not include the electronic potential. However, the curves for $V_{atom}$ include both the electronic and nuclear potentials.**

| Model | $\Psi_i$ | $\Psi_f$ | $V_i^{Cl}$ | $V_i^{SQM}$ |
|---|---|---|---|---|
| 5-Body | $\chi_i(\vec{R_i})\phi_p(\vec{r_{41}})\Phi_{He}(\vec{r_2},\vec{r_3})$ | $\chi_f(\vec{R_f})\Phi_{pb}(\vec{r_{41}},\vec{r_{31}})\xi_{He^+}(\vec{r_2})$ | $\frac{Z_\alpha Z_\alpha}{r_1} + \frac{Z_\alpha Z_e}{r_{12}} + \frac{Z_\alpha Z_e}{r_{13}}$ $+ \frac{Z_e Z_e}{r_{43}} + \frac{Z_e Z_e}{r_{42}} + \frac{Z_e Z_\alpha}{r_4}$ | $2Z_\alpha \alpha \left( e^{-2\alpha r_1} + \frac{e^{-2\alpha r_1}}{\alpha r_1} - \frac{1}{\alpha r_1} \right)$ $+ 2Z_e \alpha \left( e^{-2\alpha r_4} + \frac{e^{-2\alpha r_4}}{\alpha r_4} - \frac{1}{r_4} \right)$ |
| SBS | $\chi_i(\vec{R_i})\Phi_{He}(\vec{r_2},\vec{r_3})$ | $\chi_f(\vec{R_f})\phi_p(\vec{r_{31}})\xi_{He^+}(\vec{r_2})$ | $\frac{Z_{pi}Z_\alpha}{r_1} + \frac{Z_{pi}Z_e}{r_{12}} + \frac{Z_{pi}Z_e}{r_{13}}$ | $2\alpha Z_{pi} \left( e^{-2\alpha r_1} + \frac{e^{-2\alpha r_1}}{\alpha r_1} - \frac{1}{\alpha r_1} \right)$ $+ \frac{Z_{pi}Z_\alpha}{r_1}$ |
| DBS | $\chi_i(\vec{R_i})\phi_p(\vec{r_{41}})\Phi_{He}(\vec{r_3})$ | $\chi_f(\vec{R_f})\Phi_{pb}(\vec{r_{41}},\vec{r_{31}})$ | $\frac{Z_\alpha Z_{ti}}{r_1} + \frac{Z_\alpha Z_e}{r_{13}} + \frac{Z_e Z_e}{r_{43}}$ $+ \frac{Z_e Z_{ti}}{r_4}$ | $Z_\alpha \left( Z_{ti} e^{-2Z_{ti}r_1} + \frac{e^{-2Z_{ti}r_1}}{r_1} \right.$ $\left. + \frac{Z_{ti} - 1}{r_1} \right)$ $+ Z_e \left( Z_{ti} e^{-2Z_{ti}r_4} + \frac{e^{-2Z_{ti}r_4}}{r_4} \right.$ $\left. + \frac{Z_{ti} - 1}{r_4} \right)$ |
| DBS-Y | $\chi_i(\vec{R_i})\phi_p(\vec{r_{41}})\Phi_{He}(\vec{r_3})$ | $\chi_f(\vec{R_f})\Phi_{pb}(\vec{r_{41}},\vec{r_{31}})$ | $\frac{Z_\alpha Z_{ti}}{r_1} + \frac{Z_\alpha Z_e}{r_{13}} + \frac{Z_e Z_e}{r_{43}}$ $+ \frac{Z_e Z_{ti}}{r_4}$ | $\frac{Z_\alpha Z_{ti} e^{-\beta r_1}}{r_1} + \frac{Z_\alpha Z_e}{r_{13}} + \frac{Z_e Z_e}{r_{43}}$ $+ \frac{Z_e Z_{ti} e^{-\beta r_4}}{r_4}$ |

**Table 1 Comparison of initial and final state wave functions and perturbations for the models as described in the text.**

For a complete analysis of the models presented here, it is necessary to discuss the screening of the nuclear charge. Because all of the models contain some approximations to the two-electron bound states, some screening of the nuclear charge is possible. Thus, we have a choice in the values of $Z_{pi}$ and $Z_{ti}$, which will affect the binding energies used in the calculation. Table 2 contains a list of the values used for the results shown below. A qualitative discussion of their effects is contained in Section 3.

| Model | $Z_{pi}$ | $Z_{ti}$ | $B_{ai}$ (eV) | $B_{pi}$ (eV) | $B_{af}$ (eV) | $B_{pf}$ (eV) |
|---|---|---|---|---|---|---|
| 5-Body | NA | NA | -79.0 | -54.4 | -54.4 | -79.0 |
| SBS | 1 | NA | -79.0 | NA | -54.4 | -13.6 |
| SBS-C | 1 | NA | -79.0 | NA | -54.4 | -13.6 |
| DBS | NA | 1 | -13.6 | -54.4 | NA | -79.0 |

**Table 2 Charges and binding energies used to produce the results of Section 3. The binding energies are: $B_{ai}$ initial state target atom, $B_{pi}$ incident projectile, $B_{af}$ residual ion, $B_{pf}$ scattered projectile.**

## 3. Results

As mentioned above, we consider two collision processes with $^3$He$^+$ and $^4$He$^+$ projectiles colliding with neutral $^4$He. Results from all of the models discussed above are compared with two separate sets of experimental data in different energy regimes. We begin by comparing the results of the models to the experimental results of [17] with a $^3$He$^+$ projectile. These projectile energies range from 180 keV to 1.89 MeV for which the PWBA should be valid. Perturbative models are generally expected to work well when the perturbation parameter ($\eta = \frac{Z_p}{v_p}$) is less than 1. For the velocities presented in Fig 6, the perturbation parameters range from $\eta = 0.64$ (180 $keV$) to $\eta = 0.2$ (1.89 $MeV$).

The differential cross sections for the models with the classical perturbation all show a deep minimum. This is commonly known to be caused by a cancellation of terms in the perturbation potential [17,18,20-23]. This minimum occurs at different scattering angles for each of the models due to their different perturbations. The SBS model has the minimum at the largest scattering angle and the DBS model has it at the smallest scattering angle. The 5-Body model minimum appears between those of the SBS and DBS models. This is reasonable since the 5-Body model contains the target atom bound state of the SBS model and the projectile bound state of the DBS model. Also, the 5-Body perturbation contains terms similar to the sum of the SBS and DBS models. This minimum appears at smaller scattering angles as projectile energy increases. All of the classical perturbation models accurately predict the cross section magnitude at small scattering angles, where projectile-nuclear effects are less important. However, for large scattering angles, the 5-Body model is one to two orders of magnitude larger than the SBS or DBS models. Also, at the largest projectile energy, the 5-Body model is larger than the SBS and DBS models at all scattering angles.

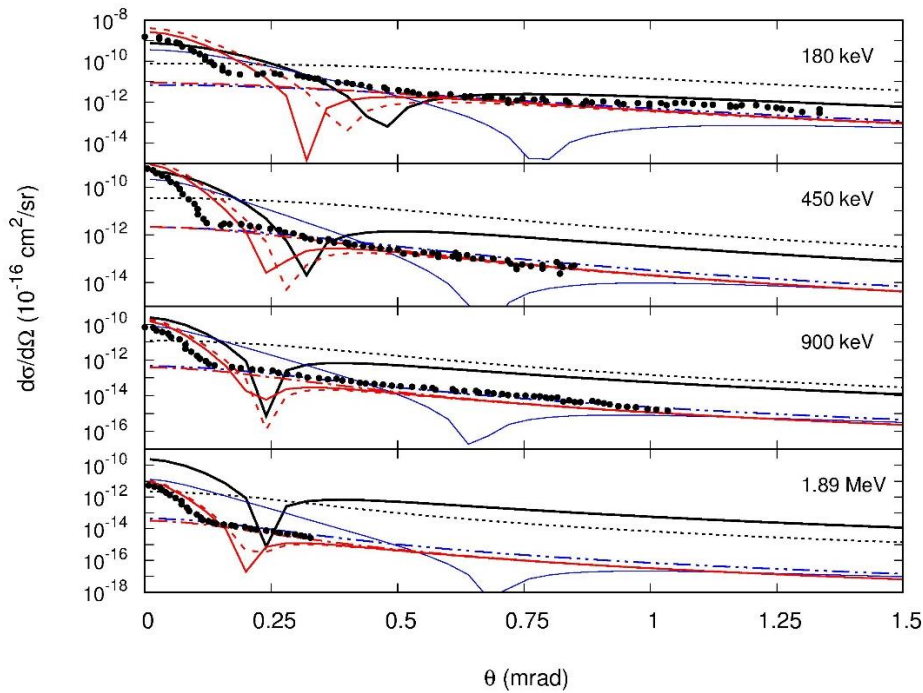

**Figure 6 Lab frame differential cross sections using the PWBA models described in Section 2. Current models are compared to the experimental results of [17].**

One might instinctively assume that the 5-Body model should be more accurate since it does not make any simplifying assumptions about the inactive electron in the collision. However, it is not clear from Fig 6 that the 5-Body model is any more accurate at predicting experiment than the SBS or DBS models. In general, the experimental data are best described by the DBS model at both large and small scattering angles, with the exception of the unphysical minimum.

Generally, the SBS model does a poor job of predicting the experimental data. The unphysical minimum occurs at a large scattering angle, and the model overestimates the cross section at small angles, while underestimating it at large angles. This indicates that projectiles with electronic structure cannot be treated as point particles using the frozen core approximation. Their electronic structure must be included in the wave function and their interaction with the target atom needs to be included in the perturbation. Given that the electronic structure of the projectile is important, it is surprising that the 5-Body model does so poorly. This is primarily a result of the location of the minimum, which, like the SBS model, occurs at too large of a scattering angle. The structure of the projectile is included in the calculation in two ways: the wave function and the perturbation. Both the 5-Body and DBS models have the same projectile bound electron wave functions, but their perturbations are different. Therefore, the poor performance of the 5-Body model seems to indicate that it is the perturbation that is more important than the wave function. This is consistent with previous work that examined the effect of the frozen core approximation in ionization collisions [15,16].

Because the deep minimum in the cross section is known to come from a cancellation of terms in the potential, we also consider the semi-quantum mechanical perturbation described above in which the atomic electrons are not considered point particles, but rather a quantum mechanical electron cloud. This SQM perturbation softens the projectile-electron interaction and removes the cancellation of terms in the potential. The effect of this SQM perturbation on the cross section is to dramatically lower the magnitude of the cross section at small scattering angles, while only minimally altering it at large scattering angles. From a classical perspective, small angle scattering occurs for large impact parameters, or when the interaction between the projectile and target is weak. Because the SQM perturbation has the effect of smearing out the electron charge cloud, it produces a less localized interaction. This then results in a decreased probability of capture at small scattering angles and a smaller cross section. At large scattering angles, the dominant interaction is between the projectile and target nucleus, which is modeled as Coulombic in both classical and SQM perturbations. Therefore, the large scattering angle cross sections are similar for both types of models. However, the SQM models generally do a poor job of predicting experiment.

To further explore the effect of softening the potential in the perturbation, Fig 6 also shows the differential cross section calculated using the DBS-Y model. In this model, the target nuclear interaction terms are softened by introducing a screening function. This screening of the nucleus does not remove the unphysical minimum in the cross section, but alters its location. In general, the DBS-Y and DBS models are very similar. Recall that as the screening parameter goes to zero, the DBS-Y model is identical to the DBS model. Results in Fig. 6 are shown for $\beta = 1$, however we examined other values of the screening parameter. As $\beta$ increased, the magnitude of the cross section did not change, but the location of the minimum moved to larger scattering angles. This trend persisted, regardless of the energy of the projectile.

The experimental data show a pronounced elbow where the slope of the cross section changes to become more gradual with increasing scattering angle. This is due to different capture mechanisms dominating the small and large angle scattering regimes. For small angle scattering, the dominant capture mechanism is momentum transfer to the electron, while for large angle scattering momentum transfer between the nuclei is dominant [24-26]. This again is consistent with the SQM results in which the models drastically underestimate the capture cross sections at small scattering angles, but accurately describe the experimental results in the large scattering angle regime.

As mentioned above, in the SBS and DBS models, the values of the projectile charge and target nuclear charge can be varied to account for screening by the inactive electron. Variation of these charges will then affect the binding energies used in the calculations. The results shown in Fig. 6 do not include any screening effects in the charges (see Table 2 for values used), however adjustment of the charges has some effect on the differential cross section. As $Z_{pi}$ increased from 1 to 1.6875, the deep minimum in the SBS model moved to larger scattering angles and the overall magnitude of the cross sections increased by about one order of magnitude. As $Z_{ti}$ increased from 1 to 1.6875, the deep minimum in the DBS model moved to smaller scattering angles and the overall magnitude of the cross sections increased by about one order of magnitude. These trends persisted regardless of projectile energy.

Figure 7 shows the differential cross section for slower $^4$He$^+$ projectiles with incident energies of 30 keV ($\eta = 1.8$) and 100 keV ($\eta = 1$). At 30 keV a PWBA model is not expected to work well, but a 100 keV projectile is at the low end of the models' applicability. The results in Fig 7 are plotted as $2\pi \sin \theta_s \, d\sigma/d\Omega$ on a linear scale. This leads to all models and experiment showing a zero at $\theta_s = 0$. The double peak structure in the experimental data can be attributed to Fraunhofer diffraction effects [1-6] with the minimum in the experiment occurring at the predicted first dark band of a single slit diffraction pattern. These features have been accurately reproduced with models based on semiclassical atomic or molecular orbital expansions [2,3,5,6]. In these models, a direct comparison between the mathematical expression for the differential cross section and that of the light intensity in classical single slit diffraction shows nearly identical functional forms [3]. It is therefore not surprising that diffraction-like oscillations in the cross section are observed. These features have not yet been observed in fully quantum mechanical models applied to faster projectiles, likely due to the fact that diffraction effects would be much more difficult to observe at larger projectile energies. In classical optics, the diffraction rings have an angular spacing of approximately $\frac{\lambda}{2R}$. At the atomic scale, projectile wavelengths range from $10^{-2}$ a.u. (100 eV proton) to $10^{-4}$ a.u. (4 MeV alpha particle) and atomic interaction ranges are on the order of 10 $a_0$. This leads to angular spacing of the diffraction peaks between 0.5 mrad and 5 μrad. Therefore, in order for current experimental techniques to observe diffraction effects, projectile energies must be in the low energy regime where semiclassical models are most appropriate.

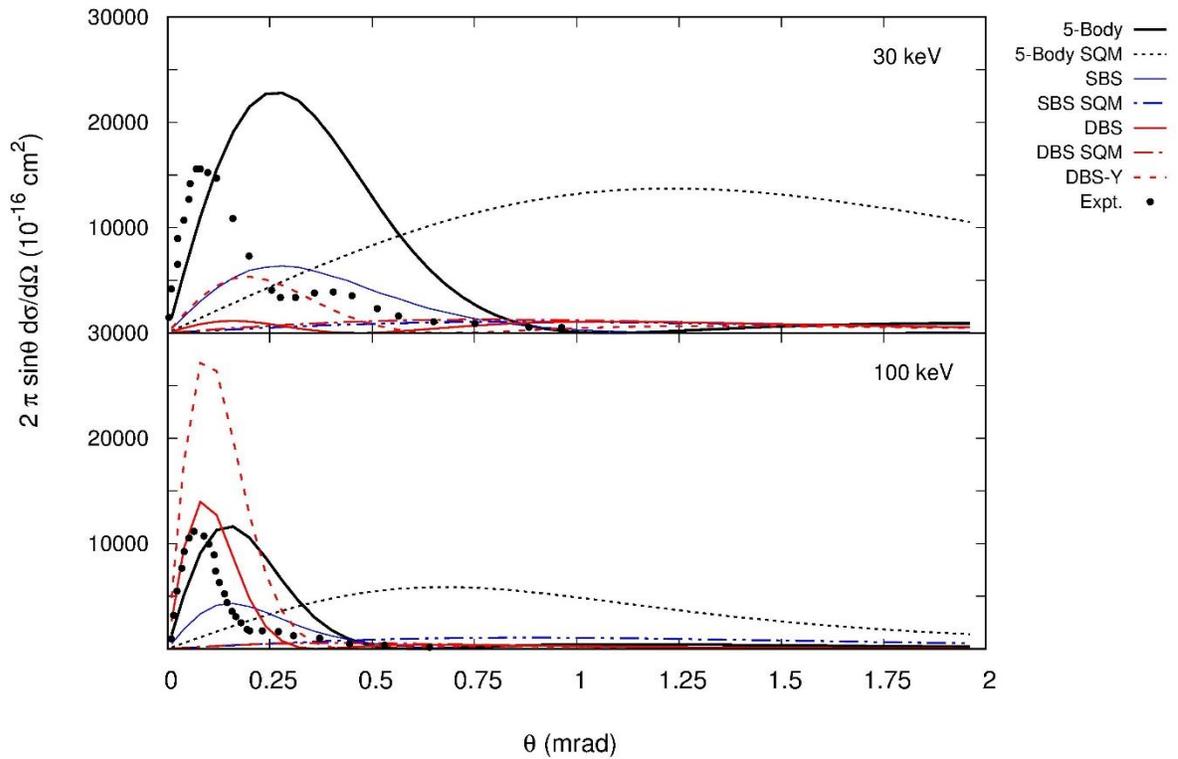

**Figure 7** Lab frame differential cross sections using the PWBA models described in Section 2. Current models are compared to the experimental results of [1] and all theoretical results have been divided by 10.

From Fig 7 it can be seen that all of the PWBA models using the classical perturbation show a two peak structure, although the second peak is not readily noticeable on a linear scale. This peak structure is not associated with diffraction effects, but is due to the unphysical minimum in the cross section.

As with the higher energy results, none of the models accurately predict experiment and in fact they all overestimate the experimental data by approximately one order of magnitude. This overestimation is well-known for the PWBA, particularly at lower energies. The DBS model again shows the best agreement at small scattering angles, while the SBS and 5-Body models predict too broad of a dominant peak located at too large of a scattering angle. This broadening and shifting of the dominant peak to larger scattering angles indicates that the models overestimate the probability for capture in grazing collisions. As noted with the higher projectile energies, this is likely due to the projectile-nucleus interaction being too strong in the 5-Body and SBS models.

A striking difference is observed in the cross sections for the models using the classical perturbation compared to those of the SQM perturbation. The SQM models predict a very large broad cross section that in no way resembles the experimental data, even at large scattering angles where for higher projectile energies these models showed good agreement with experiment. The forward scattering capture cross section is significantly underestimated, which is again due to the projectile-electron interaction being spatially smeared out causing a weaker interaction. Overall, Fig 7 clearly shows the inadequacy of the PWBA models at low projectile energies and in particular the importance of the projectile-electron interaction for predicting capture at small scattering angle.

## 4. Conclusion

We have presented results from three different plane wave Born approximation models using two types of perturbations. In all models, the incident and scattered projectiles were treated as plane waves and any two-particle bound states were approximated as product wave functions within the independent particle model. While these are very simple models, they show qualitatively some important physical effects. In the 5-Body model, all particles in the collision system were explicitly included in the calculation, while in the Single Bound State and Double Bound State models, either the incident projectile or inactive target atom electron was neglected, respectively. The differences between the 5-Body and SBS or DBS models showed the effect of the frozen core approximation. When a classical perturbation was used, all models showed an unphysical minimum due to a cancellation in the terms in the perturbation. Comparison of the 5-Body, SBS, and DBS models revealed that the neglect of inactive electrons in the perturbation had the most significant effect on the shape of the cross section.

The deep minimum in the cross section with the classical perturbation prompted us to develop a semi-quantum mechanical perturbation potential in which the atomic electrons are modeled as a quantum mechanical electron cloud rather than point particles. Use of the SQM perturbation removed the unphysical minimum, but it also lowered the magnitude of the cross section at small scattering angles. This was attributed to the smearing out of the electron probability density, which reduces the small angle capture probability. We also presented results from the DBS-Y model, which introduced screening of the target nucleus. Like the classical perturbation models, the DBS-Y model also predicts a deep minimum in the cross section, showing that there is still a cancellation of terms in the potential, despite the nuclear screening.

While the results presented here come from rudimentary models, they demonstrate important physical concepts involving diffraction effects and the role of inactive electrons in the single electron capture process and the classical vs. quantum mechanical description of the projectile-electron interaction. We anticipate that more sophisticated quantum mechanical models will be able to more accurately predict experimental results on a quantitative level.

## Acknowledgements

We gratefully acknowledge the support of the NSF under Grant Nos. PHY-1505217 and PHY-1838550.